\documentclass[12pt]{article}

\input epsf.tex
\usepackage{amsmath,bbm}
\usepackage{amssymb}
\usepackage{graphicx}
\usepackage[usenames]{color}

\begin{document}
\setlength{\textheight}{23cm}
\setlength{\topmargin}{-1cm}
\renewcommand{\thefootnote}{\fnsymbol{footnote}}
\font\vmath=msbm10 at 12pt
\newcommand{\vmi}[1]{\mathbbm{#1}}
\def\sp{\phantom{a}}

\begin{titlepage}
\font\csc=cmcsc10 scaled\magstep1
{\baselineskip=14pt
 \rightline{
 \vbox{\hbox{RIKEN-TH-22}}}}
{\baselineskip=14pt
\rightline{
 \vbox{\hbox{hep-th/0404206}
    }}}

 \renewcommand{\thefootnote}{\fnsymbol{footnote}}
 
     \begin{Large}
       \vspace{2cm}
       \begin{center}
         {Path integral formulation of noncommutative superspace in IKKT
	matrix model}      \\
       \end{center}
    \end{Large}

  \vspace{1cm}

\begin{center}
           SHIBUSA, Yuuichirou  \footnote
           {
e-mail address : shibusa@riken.jp} 
          
       \it Theoretical Physics Laboratory\\
          The Institute of Physical and Chemical Research (RIKEN)\\
          Wako 2-1, Saitama 351-0198, Japan

       \end{center}

\vfill

\begin{abstract}
We propose a physical interpretation of our novel fermionic solution 
for the IKKT matrix model which obtained in our previous 
paper \cite{Shibusa:2003dg}. We extend the
 configuration space of bosonic field to supernumbers space and obtain 
the noncommutative parameter which is not bi-grassmann but  an ordinary 
number. This establishes the connection between Seiberg's noncommutative  
superspace and our solution of the IKKT matrix model.

\end{abstract}

\vfill
\end{titlepage}
\vfil\eject

\baselineskip=16pt
\setcounter{footnote}{0}
\renewcommand{\thefootnote}{\arabic{footnote}}
%\pagestyle{empty}
%\tableofcontents
%\newpage

\section{Introduction}

For several years, the noncommutative space in string theory has been 
studied actively 
\cite{Connes:1997cr, Seiberg:1999vs, Aoki:1999vr, Douglas:2001ba}.
At first, only the noncommutative structure of bosonic
coordinates was studied. Recently, this concepts is extended to the superspace 
\cite{Seiberg:2003yz}. Many attempts to study this new noncommutativity 
have been made. For example, the solutions that represents noncommutative
superspace were introduced in the context of super-Lie 
matrix model
\cite{Hatsuda:2003ry,Iso:2003zb,Iso:2003fp,Park:2003ku,Kurkcuoglu:2003ke,
Ivanov:2003qq}.
And many attempts have been also made in the context of field theory,
quantum mechanics, string theory \cite{Klemm:2001yu,deBoer:2003dn,
Ooguri:2003qp,Ooguri:2003tt,Kawai:2003yf,Britto:2003aj,Terashima:2003ri,Ferrara:2003xy,Britto:2003aj2,Britto:2003kg,Lunin:2003bm,Berenstein:2003sr,Sako:2003jx,Chandrasekhar:2003uq,Hatsuda:2003wt,Imaanpur:2003ig,Britto:2003uv,Kawai:2003dw,Grisaru:2004qw,Inami:2004sq,Morita:2004yr}.
In this paper, we would like to shed the lights on this noncommutativity 
from the view point of the IKKT matrix model \cite{Ishibashi:1996xs}.

While one can investigate the IKKT matrix model
by the expansion around the commutative (diagonal) background 
\cite{Aoki:1998vn}, it is also 
interesting to start with the following solution:
\begin{equation}
[X^{a},X^{b}]=-i C^{ab} { \vmi{ 1}}_{N}, \theta^{\alpha}=0, 
\label{pqsol}
\end{equation}
where $C^{ab}$ is an antisymmetric constant. The result is
noncommutative 
Yang-Mills theory \cite{Aoki:1999vr}. $X_{a}$'s which obey 
(\ref{pqsol}) serve as noncommutative coordinates. 
In the string picture, this corresponds to the appearance of 
noncommutative geometry in a constant B field background \cite{Seiberg:1999vs}.

On the other hand, for the backgrounds of graviphoton fields, the end points of
open strings have the structure of Seiberg's noncommutative superspace 
\cite{Berkovits:2003kj}. 
Should the IKKT matrix model contains the whole string theory, it must have 
the solution which corresponds to this noncommutativity. We have provided a 
candidate for this vacuum in our previous paper \cite{Shibusa:2003dg}.
However, in this solution the noncommutative parameter is not an ordinary 
number but bi-grassmann. In the string theory, this parameter is an ordinary
number which depends on $\alpha'$ and graviphoton fieldstrength.
Thus we need to clarify the physical interpretation of 
this bi-grassmann structure, in order to claim that this solution of 
the IKKT matrix model
corresponds to the string theory in the backgrounds of graviphoton
field.

In this paper, we interpret this noncommutative parameter as vacuum
expectation value of this solution. 
In the process, we will deform the path of the integral to a special 
path in the supernumber configuration space. Our computation involves
new techniques, which we will explain in detail.

This paper is organized as follows. In section 2, we review the 
concept of supernumber which is necessary to define path
integrals over our new configuration space. In section 3, we introduce 
novel method to compute path integrals. And we compute the 
commutation relation of coordinates of superspace: $x^a, \theta^{\alpha},
\bar{\theta}^{\dot{\alpha}}$. In section 4, we discuss our method and
provide the meaning of it.

\section{Supernumbers}

In this section, we introduce the concepts of supernumbers and 
integrals over supernumbers \cite{DeWitt:cy}.

Let $\zeta^i, a=1 \cdots M$, be a set of generators for an algebra, 
which satisfy the following anticommutation relation:
\begin{eqnarray}
 \zeta^i\zeta^j=-\zeta^j\zeta^i \sp \sp \sp \mbox{for all} \sp i,j.
\end{eqnarray}
This algebra is called `Grassmann algebra' and is denoted by
$\Lambda_M$. The elements $1,\zeta^i,\zeta^i\zeta^j,\cdots$ form the basis of
$\Lambda_M$.

The elements of $\Lambda_M$ will be called supernumbers. Every
supernumber can be expressed in the form
\begin{eqnarray}
z=z_B+z_S =z_B+\sum_{m=1}^M
 \frac{1}{m!}c_{i_1\cdots i_m}\zeta^{i_m}\cdots \zeta^{i_1}.
\end{eqnarray}
This $z_B$ is an ordinary number and is called body. $z_S$ is called soul.
We can divide supernumbers into two sectors: one whose basis consists of
even number of $\zeta^i$'s `$z_c$' and that of odd number of
$\zeta^i$'s `$z_a$'.
\begin{eqnarray}
z&=&z_c+z_a, \\
z_c&=&z_B+\sum_{m=1}^{[\frac{M}{2}]}\frac{1}{(2m)!}c_{i_1 \cdots
 i_{2m}}\zeta^{i_{2m}}\cdots \zeta^{i_1}, \\
z_a&=&\sum_{m=1}^{[\frac{M}{2}]}\frac{1}{(2m-1)!}c_{i_1 \cdots
 i_{2m-1}}\zeta^{i_{2m-1}}\cdots \zeta^{i_1}.
\end{eqnarray}
The supernumber $z$ which is $z_c=0$ is called a-number and the
supernumber $z$ which is $z_a=0$ is called c-number.
The set of c-numbers is denoted by $\Re_c$ and the set of a-numbers is
denoted by $\Re_a$.

We can define a function over $\Lambda_M$ as follows,
\begin{eqnarray}
f(z)\equiv \sum_{m=0}^M\frac{1}{m!}f^{(m)}(z_B)(z_S)^m,
\end{eqnarray}
where $f^{(m)}(z_B)$ are m-th derivatives with respect to $x$ of usual
function $f(x)$.
Also we can define the integrals of $f(z)$ over $\Re_a$ and $\Re_c$.

The integrals over $\Re_a$ are called Berezin's integral.
The functions of $z_a$ can be denoted as $f(z_a)=a+bz_a$, hence we define
the integrals over $\Re_a$ as follows,
\begin{eqnarray}
\int dz_a&=&0, \\
\int dz_a \sp z_a&=&1,\\
\int dz_a \sp af(z_a)+bg(z_a)&=&a\int dz_a \sp f(z_a)+b\int dz_a \sp g(z_a),
\end{eqnarray}
where $a,b$ are constant supernumbers which do not depend on $z_a$.

The integrals over path $C$ in $\Re_c$ are defined as follows. 
Consider a path $C$ as $z_S=z_S(z_B)$, which starts for $a$ and ends at
$b$. We define the integral for $C$ as follows,
\begin{eqnarray}
\int_a^b dz f(z)=\sum_{m=0}^M
 \frac{1}{m!}\int_{a_B}^{b_B}dz_B(1+z_S'(z_B)) f^{(m)}(z_B)z_S^m(z_B).
\end{eqnarray}
From this definition, the integral depends only on the starting point $a$ and
the endpoint $b$.

When we study the IKKT matrix model, we consider $\Re$ as a configuration
space of bosonic fields which are components of Lie algebra $X^a$ and 
grassmann number as a configuration space of fermionic fields which are
components of Lie algebra $\theta,\bar{\theta}$.

In the following, we propose to extend the configuration space of
$X^a$ from $\Re$ to $\Re_c$, and perform path integrals over $\Re_c$.
As above, the integrals do not vary by a deformation of the path, hence we can
expect that the theory does not change.

\section{Noncommutative superspace}

In our previous paper \cite{Shibusa:2003dg}, we obtain the following new
solution of IKKT matrix model:
\begin{eqnarray}
\label{solTh}
\sum_{AB}\theta^{\alpha A}\theta^{\beta B}f_{AB0}&=& \tilde{C}^{\alpha \beta},\\
%f_{AB} is replaced with f_{AB0}
\label{solbth}
\bar{\theta}^{\dot{\alpha}\sp  0}&=& \bar{\theta}^{\dot{\alpha}},\\
\label{solX}
X^{aA}&=&-i\theta^{\alpha A}(\sigma^a)_{\alpha \dot{\beta}}\bar{\theta}^{ \dot{\beta}0},  \\
\mbox{Others}&=& 0. \label{solO}
\end{eqnarray}

Here we use unknown characters which are defined as follows and the
spinor notations of \cite{Wess:cp}. 

Let us denote $U(N)$ generators as $T^{\hat{A}}$ and choose a integer $n$ which is large enough, but much smaller than $N$ so that $N/n >>1$.

\begin{eqnarray}
\hat{A}&=&0,1,2,\cdots, N^2-1, \nonumber \\
A&=&1,2, \dots , 2n, \nonumber \\
T^A&=& Q_1,P_1,Q_2,P_2,\dots , Q_{n}, P_{n},
\end{eqnarray}
where $Q_{k}$'s and $P_{k}$'s are $N/n \times N/n$ matrices and  they satisfy
\begin{equation}
[Q_{j}, P_{k}] = i \delta_{jk}.
\end{equation}
We also express the identity by $T^0$, that is,
\begin{equation}
T^0= \vmi{ 1} _{N}.
\end{equation}
It follows that the only nontrivial structure constants
$f_{\hat{A}\hat{B}\hat{C}}$ among the generators $T^A$ are
\begin{eqnarray}
%f_{AB}&\equiv& f_{AB0}, \nonumber \\
f_{AB0}&=&
\left(
\begin{array}{cccccc}
 0  & i & 0 & 0  & 0 & 0\\
 -i  & 0 & 0 & 0 & 0 & 0\\
 0  &  0 & 0 & i & 0 & 0\\
 0  &  0 & -i & 0 & 0 & 0 \\
 0 & 0 & 0 & 0 & \cdot &   \\
 0 & 0 & 0 & 0 &  & \cdot
\end{array}
\right). 
\end{eqnarray}

(\ref{solTh})-(\ref{solX}) are solutions of 4-dimensional IKKT matrix model,  
\begin{eqnarray}
\label{action}
S&=&Tr(\frac14 [X^a,X^b]^2- \frac{1}{2}\theta\sigma^a [\bar{\theta},X_a]),
\end{eqnarray}
which yields the following equations of motion, 
\begin{eqnarray}
\label{motionX}
[X^b,[X^a,X_b]]-\frac{1}{2}
\{ \theta ,\sigma^a\bar{\theta} \} &=& 0, \\
\label{motionTh}
[X_a, (\theta\sigma^a)_{\dot{\alpha}}]&=&0, \\
\label{motionbTh}
[(\sigma^a \bar{\theta})_{\alpha}, X_a]&=&0.
\end{eqnarray}

The special feature of this solution is that they obey the following algebra
of the Seiberg's noncommutative algebra \cite{Seiberg:2003yz}
\begin{eqnarray}
\label{ferminoncom0}
&&\{\theta^{\alpha},\theta^{\beta}\}=\tilde{C}^{\alpha\beta},\\
\label{boseferminoncom0}
&&[X^a,\theta^{\alpha}]=i\tilde{C}^{\alpha\beta}\sigma^a_{\beta \dot{\alpha}}
\bar{\theta}^{\dot{\alpha}},\\
\label{bosenoncom0}
&&[X^a,X^b]=(\bar{\theta})^2\tilde{C}^{ab},  \\
&&\{{\bar \theta}^{\dot \alpha},{\bar \theta}^{\dot \beta}\}=\{{\bar \theta}^{\dot \alpha},{ \theta}^{\beta}\} = [{\bar \theta}^{\dot \alpha}, X^{a}]=0, \label{fermibar0}
\end{eqnarray}
where\begin{eqnarray}
a,b &=& 1 \dots 4, \sp \sp \alpha,\dot{\alpha}=1 ,2, \nonumber \\
\label{Cdef01}
\tilde{C}^{ab}&\equiv&\tilde{C}^{\alpha\beta}(-\sigma^{ab} \epsilon)_{\alpha\beta}, \\
\label{Cdef02}
\tilde{C}^{\alpha \beta}&=& (\epsilon\sigma^{ab})^{\alpha \beta}\tilde{C}_{ab}.
\end{eqnarray}

This algebra is the same as the Seiberg's noncommutative algebra \cite{Seiberg:2003yz},
\begin{eqnarray}
\label{ferminoncom}
&&\{\theta^{\alpha},\theta^{\beta}\}=C^{\alpha\beta},\\
\label{boseferminoncom}
&&[X^a,\theta^{\alpha}]=iC^{\alpha\beta}\sigma^a_{\beta \dot{\alpha}}
\bar{\theta}^{\dot{\alpha}},\\
\label{bosenoncom}
&&[X^a,X^b]=(\bar{\theta})^2C^{ab},  \\
&&\{{\bar \theta}^{\dot \alpha},{\bar \theta}^{\dot \beta}\}=\{{\bar \theta}^{\dot \alpha},{ \theta}^{\beta}\} = [{\bar \theta}^{\dot \alpha}, X^{a}]=0, \label{fermibar}
\end{eqnarray}
where\begin{eqnarray}
a,b &=& 1 \dots 4, \sp \sp \alpha,\dot{\alpha}=1 ,2, \nonumber \\
\label{Cdef1}
C^{ab}&\equiv&C^{\alpha\beta}(-\sigma^{ab} \epsilon)_{\alpha\beta}, \\
\label{Cdef2}
C^{\alpha \beta}&=& (\epsilon\sigma^{ab})^{\alpha \beta}C_{ab}.
\end{eqnarray}
We might identify this solution with Seiberg's noncommutative superspace
recognizing that $\tilde{C}^{\alpha \beta}$ corresponds to
$C^{\alpha \beta}$. However, the noncommutative parameter
$\tilde{C}^{\alpha \beta}$ is bi-grassmann (\ref{solTh}) while 
$C^{\alpha \beta}$ is an ordinary number which depends on $\alpha'$ and
graviphoton fieldstrength \cite{Berkovits:2003kj}.
Thus physical interpretation of the solution would not be trivial.

In the following, we present a physical interpretation of this solution 
by showing that Seiberg's noncommutative algebra 
(\ref{ferminoncom})-(\ref{fermibar}) can be rather realized as vacuum
expectation values of the algebra for this solution.

\begin{eqnarray}
\label{expferminoncom}
\langle Tr(\{\theta^{\alpha},\theta^{\beta}\})\rangle&=&C^{\alpha\beta},\\
\label{expboseferminoncom}
\langle Tr([X^a,\theta^{\alpha}])\rangle &=&iC^{\alpha\beta}\sigma^a_{\beta 
\dot{\alpha}} \bar{\theta}^{\dot{\alpha}},\\
\label{expbosenoncom}
\langle Tr([X^a,X^b])\rangle&=&(\bar{\theta})^2C^{ab}, \\
\mbox{Others}&=&0.  
\label{expother}
\end{eqnarray}
Here $C_{\alpha \beta}, C_{ab}$ are ordinary numbers which will be
obtained below. Thus we can exactly
identify these parameter as Seiberg's noncommutative parameters. 

Let us proceed to the actual computation in the path integral
formulation. What we need to compute are various vacuum expectation values,
\begin{eqnarray}
\label{exp}
\langle \mathcal{O} \rangle &=& \int \mathcal{D}X \mathcal{D}\theta
\mathcal{D}\bar{\theta}\sp \mathcal{O}e^{-S}.
\end{eqnarray}
The measure of this integration is given by
\begin{eqnarray}
\mathcal{D}X^a\mathcal{D}\theta\mathcal{D}\bar{\theta}=\prod_{\tilde{A}=1}^{\tilde{A}=N^2-1}dX^a_{\tilde{A}}d^2\theta_{\tilde{A}}d^2\bar{\theta}_{\tilde{A}}.
\end{eqnarray}
We decompose the components of matrix $X^a, \theta, \bar{\theta}$, as
\begin{eqnarray}
X^a &=& \sum_{\tilde{A}=1}^{\tilde{A}=N^2-1}
 T^{\tilde{A}}X^a_{\tilde{A}}, \\
\theta &=& \sum_{\tilde{A}=1}^{\tilde{A}=N^2-1}
 T^{\tilde{A}}\theta_{\tilde{A}}, \\
\bar{\theta} &=& \sum_{\tilde{A}=1}^{\tilde{A}=N^2-1}
 T^{\tilde{A}}\bar{\theta}_{\tilde{A}}.
\end{eqnarray}
As above, the gauge group is SU(N).

Because this ``vacuum'' (\ref{solTh})-(\ref{solX}) is not the ordinary vacuum 
but Seiberg's noncommutative superspace vacuum, 
we need to perform the following ``tree'' approximation.
\begin{figure}[h]
\leavevmode\epsfxsize=0.95\columnwidth \epsfbox{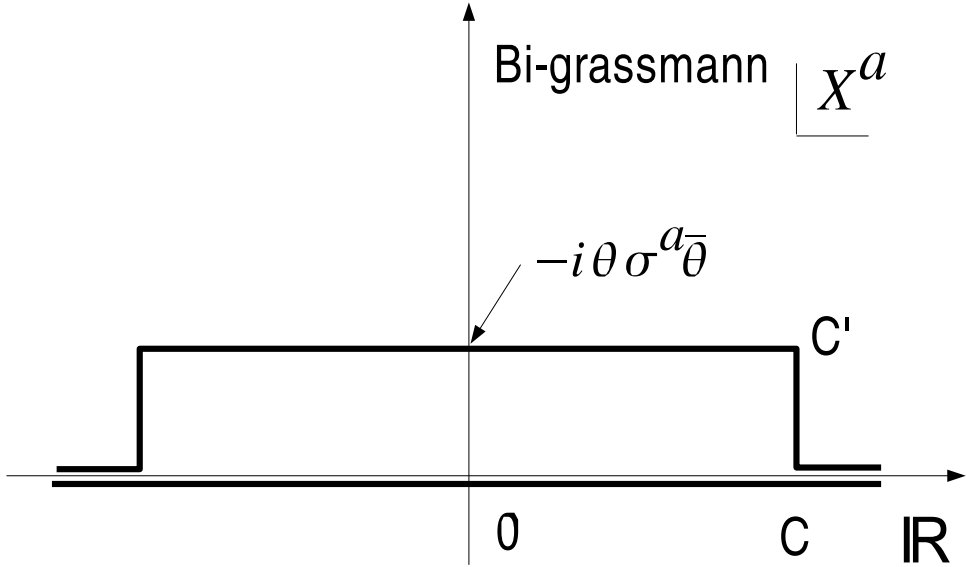}
\caption{Deformation of integral path}
\label{fig:path}
\end{figure}
                                                                                
We deform the integral path of $X^a$ from $C$ to $C'$ (Fig.~\ref{fig:path}).
Because the integrals do not depend on the path, we can perform the
integration over the path $C'$. In other words, we change the variables
from $X^a$ to $\tilde{X^a}+ -i\theta\sigma^a \bar{\theta}$. 
Because $\tilde{X^a} \in \Re$ is an ordinary number,
we can adopt the value at $\tilde{X^a}=0$ as a first order approximation to
the integrals over $\tilde{X^a}$. Thus we obtain

\begin{eqnarray}
\label{tree}
\langle \mathcal{O} \rangle &=& \int \mathcal{D}X \mathcal{D}\theta
\mathcal{D}\bar{\theta}\sp \mathcal{O}e^{-S} \nonumber \\
 &\approx& \int \mathcal{D}\theta \mathcal{D}\bar{\theta}\sp \mathcal{O}e^{-S} 
 \big|_{X^a=-i \theta\sigma^a \bar{\theta}}. \\
&\mbox{``tree''}& \nonumber
\end{eqnarray}

We call this approximation as ``tree''. We do not insist that we accept 
this value as the first order approximation of the entire IKKT matrix
model. This approximation is effective only in the vicinity of
this Seiberg's noncommutative superspace vacuum $\tilde{X^a}=0$. Even on this
path $C'$, there are other vacua, for example, $\tilde{X^a}=\mbox{usual
noncommutative plane}$.  

Action $S[X^a=-i\theta\sigma^a \bar{\theta}](\equiv \tilde{S})$ is 
fourth order of 
$\theta,\bar{\theta}$ and has zero modes (\ref{solTh}),(\ref{solbth}).
Strictly speaking, we are not handling the gauge group U(N) but SU(N),
hence $\bar{\theta}=\bar{\theta}_0 \vmi{ 1} _{N}$ is not contained in
the model. However, because we would like to treat this vacuum 
which is asymmetry between $\theta$ and
$\bar{\theta}$, we adopt the following technique. 

We extend the gauge group SU(N) to U(N) only on $\bar{\theta}$.
\begin{eqnarray}
\mathcal{D}\theta \mathcal{D}\bar{\theta}&\rightarrow&\mathcal{D}\theta
 \mathcal{D}\bar{\theta}d^2\bar{\theta}_0, \\
\bar{\theta} &\rightarrow& \bar{\theta}=\sum_{\tilde{A}=0}^{\tilde{A}=N^2-1}
 T^{\tilde{A}}\bar{\theta}_{\tilde{A}}.
\end{eqnarray}
This model contains the zero modes (\ref{solTh}),(\ref{solbth}) but does
not contain the zero modes which are obtained by exchanging  
$\theta \leftrightarrow \bar{\theta}$. We define the expectation 
value of this model as
$\langle \mathcal{O} \rangle_{\rm{deformed}}$.

Similarly to the case of the instanton backgrounds, 
when we treat a theory which
has fermionic zero modes, the expectation value of operators $\mathcal{O}$ 
is zero if $\mathcal{O}$ does not contain all 
the zero modes \cite{Weinberg:kr},

\begin{eqnarray}
\label{exp1}
\langle 1 \rangle_{\rm{deformed}}&=& \int \mathcal{D}\theta
 \mathcal{D}\bar{\theta}d^2\bar{\theta}_0 e^{-\tilde{S}}=0. 
\end{eqnarray}
Because we have fermionic zero modes (\ref{solTh}) and
(\ref{solbth}) in this model, we must compute expectation value of 
the operator which contains all the fermionic zero modes $\bar{\theta}_0^2$, 
$\theta_{sol}^2$, in order to obtain non-zero expectation value, 
\begin{eqnarray}
\langle \bar{\theta}^2_0\theta_{sol}^2 \rangle_{\rm{deformed}} &=& \int \mathcal{D}\theta
 \mathcal{D}\bar{\theta}d^2\bar{\theta}_0 \bar{\theta}^2_0\theta_{sol}^2
 e^{-\tilde{S}}\ne 0.
\end{eqnarray}
In particular, we can compute the following expectation value,
\begin{eqnarray}
\langle \bar{\theta}^2_0 Tr(\{\theta^{\alpha},\theta^{\beta}\}) \rangle_{\rm{deformed}} &=& \int \mathcal{D}\theta
 \mathcal{D}\bar{\theta}d^2\bar{\theta}_0 \bar{\theta}^2_0Tr(\{\theta^{\alpha},\theta^{\beta}\})e^{-\tilde{S}}=C^{\alpha\beta}.
\end{eqnarray}
The zero modes of $\theta$ are contained on the left hand side of the 
above equation and make non-zero contribution to 
$Tr(\{\theta^{\alpha},\theta^{\beta}\})$, hence the expectation value is 
non-zero. Here $C^{\alpha \beta}$ is an ordinary number which depends on
$f_{AB}$, $\sigma^a_{\alpha\dot{\alpha}}$'s and 
the way for $N,n \to \infty$. Recalling the feature of Berezin integral,
\begin{eqnarray}
&&\int d\psi \psi f = g \nonumber \\
&\Rightarrow& f=g,
\end{eqnarray}
we obtain actually
\begin{eqnarray}
\langle \bar{\theta}^2_0 Tr(\{\theta^{\alpha},\theta^{\beta}\}) \rangle_{\rm{deformed}} &=& \int \mathcal{D}\theta
 \mathcal{D}\bar{\theta}d^2\bar{\theta}_0
 \bar{\theta}^2_0Tr(\{\theta^{\alpha},\theta^{\beta}\})e^{-\tilde{S}}
 \nonumber \\
&=& \int d^2\bar{\theta}_0\bar{\theta}^2_0 \mathcal{A}=C^{\alpha \beta}, \\
\mathcal{A}&=& \int \mathcal{D}\theta
 \mathcal{D}\bar{\theta}
 Tr(\{\theta^{\alpha},\theta^{\beta}\})e^{-\tilde{S}}
 \nonumber \\
&=& \langle Tr(\{\theta^{\alpha},\theta^{\beta}\}) \rangle =C^{\alpha \beta}.
\label{expferminoncom2}
\end{eqnarray}
We can compute (\ref{expferminoncom2}) in the SU(N) theory. 
Because $C^{\alpha \beta}$ is an ordinary number, 
we obtain noncommutative parameter which is not bi-grassmann but an ordinary number.

In the same way,
\begin{eqnarray}
\langle Tr ([X^a , X^b]) \rangle_{\rm{deformed}}&=& \langle Tr ([
-i\theta\sigma^a\bar{\theta}, -i\theta\sigma^a\bar{\theta}])
\rangle_{\rm{deformed}} \nonumber \\
&=& \int d^2\bar{\theta}_0 \mathcal{B}=C^{ab}, \\
\mathcal{B}&=& \langle Tr ([X^a , X^b]) \rangle = \bar{\theta}^2_0C^{ab}.
\label{expbosenoncom2}
\end{eqnarray} 
Here $C^{ab}\equiv C^{\alpha\beta}(-\sigma^{ab} \epsilon)_{\alpha\beta},
C^{\alpha \beta}=(\epsilon\sigma^{ab})^{\alpha \beta}C_{ab}$ which
are consistent with (\ref{Cdef1}),(\ref{Cdef2}).
Similarly, 
\begin{eqnarray}
\langle \bar{\theta}_0^{\dot{\alpha}}Tr ([X^a,\theta^{\alpha}]) \rangle_{\rm{deformed}} &=& 
\langle \bar{\theta}_0^{\dot{\alpha}} Tr
([-i\theta\sigma^a\bar{\theta},\theta^{\alpha}])
\rangle_{\rm{deformed}} \nonumber \\ 
&=& \int d^2\bar{\theta}^2_0
 \bar{\theta}_0^{\dot{\alpha}}\mathcal{C}=\frac{-i}{2}C^{\alpha\beta}(\sigma^a\epsilon)^{\sp \dot{\alpha}}_{\beta},\\
\mathcal{C}&=&\langle Tr ([X^a,\theta^{\alpha}]) \rangle = iC^{\alpha\beta}\sigma^a_{\beta \dot{\alpha}}
\bar{\theta}^{\dot{\alpha}}, \\
\label{expboseferminoncom2}
\langle Tr (\{\bar{\theta}^{\dot{\alpha}},\bar{\theta}^{\dot{\beta}}\}) \rangle &=&
\langle \bar{\theta}^2_0 Tr
(\{\bar{\theta}^{\dot{\alpha}},\bar{\theta}^{\dot{\beta}}\})
\rangle_{\rm{deformed}} \nonumber \\ 
&=& 0, \\
\mbox{Others}&=& 0.
\end{eqnarray}
Thus we obtain the full algebra (\ref{expferminoncom})-(\ref{expother}).

In string theory, the noncommutative parameter $C^{\alpha \beta}$ is 
a free parameter which depends on the backgrounds of graviphoton fields. 
Thus we need to adjust the way for $N, n \to \infty$ in SU(N) theory in 
order to obtain the arbitrary parameter $C^{\alpha \beta}$. 

\section{Discussions}
In this paper we propose a physical interpretation of our novel
fermionic solution. Seiberg's noncommutative superspace algebra is
realized as vacuum expectation values of the algebra for this solution. 
We obtain an ordinary noncommutative parameter by path integral formulation.

We adopt a special method in this paper to perform the calculation. 
The point is asymmetrical treatment of $\theta$ and $\bar{\theta}$. 
We think that this asymmetrical procedure is necessary to extract 
the information from Seiberg's noncommutative superspace vacuum. 
In the finite $N$, the asymmetrical feature does not show. 
In the true $N \to \infty$ limit, this asymmetrical feature emerges from 
the measure. 
In the finite $N$, it seems that 
\begin{eqnarray}
\langle 1 \rangle \ne 0,
\end{eqnarray}
in contract to (\ref{exp1}). 
But this contribution originates in other vacua. 
If we would like to obtain the information of Seiberg's 
noncommutative superspace
vacuum, we must take the $N \to \infty$ limit as asymmetrical 
feature appears. In this limit, the contributions from other vacua will 
disappear. 

There is another subtlety. In the true $N \to \infty$ limit,  
\begin{eqnarray}
Tr(\theta[X^a, \sigma_a \bar{\theta}])\ne Tr([\theta,X^a]\sigma_a
\bar{\theta}).
\end{eqnarray}
If we would like to obtain symmetric vacuum between 
$\bar{\theta}$ and $\theta$, 
we probably must rather treat the following action
\begin{eqnarray} 
\label{action2}
S&=&STr(\frac14 [X^a,X^b]^2- \frac{1}{2}\theta\sigma^a [\bar{\theta},X_a]),
\end{eqnarray} 
where $STr$ means symmetric trace.

We focused on the IKKT model in this paper. 
However, the technique that we adopted here, may be also applicable
to other model which has fermionic zero modes, especially in the case that
there are nontrivial relations between bosonic fields and fermionic fields.
When bosonic field $\Phi(x)$ has none-zero expectation value $\Phi_{cl}$, 
we expand field around $\Phi_{cl}$ as
$\Phi(x)=\tilde{\Phi}(x)+\Phi_{cl}$. And we treat new field
$\tilde{\Phi}(x)$ as quantum field. However, in the case that we have a 
fermionic zero mode $\psi_0$, it is a mistake if we expand the fermionic
field around $\psi_0$. 
Extending configuration space to supernumbers space may be an effective
method to wider problems.

\begin{center} \begin{large}
Acknowledgments
\end{large} \end{center}
We would like to dedicate this paper to the memory of GORO. And  
we would like to thank T. Aoyama, N. Maru, T. Matsuo, T. Tada, and 
M. Tachibana for fruitful discussions.

\end{document}